\RequirePackage{lineno}

\documentclass[aps, 10.5pt, superscriptaddress, twocolumn, nofootinbib]{revtex4-2}
\usepackage[UKenglish]{babel}
\usepackage{latexsym}
\usepackage[pdftex, colorlinks=true, linkcolor=red, citecolor=blue, urlcolor=magenta, pdfstartview=FitH]{hyperref}
\usepackage{graphicx}	
\usepackage{amssymb}
\usepackage{amsmath}
\usepackage{xcolor}
\usepackage[normalem]{ulem}

\pdfoutput=1

\usepackage{bm}
\usepackage{verbatim}
\usepackage{amsmath}
\usepackage{amssymb}
\usepackage[utf8]{inputenc}
\usepackage{nicefrac}
\usepackage{braket}
\usepackage{pdfcomment}

\usepackage{pbox}
\usepackage{makecell}
\usepackage{hhline}
\usepackage{multirow}
\usepackage{tabularx}
\usepackage{soul}
\DeclareUnicodeCharacter{2212}{\textendash}

\begin{document}
%############################## TITLE #########################################
\title{Robust trapping of 2D excitons in an engineered 1D potential from proximal ferroelectric domain walls.}
%##############################################################################
%
%############################ AUTHORS #########################################

%
\author{P. Soubelet}\email{pedro.soubelet@wsi.tum.de}
\affiliation{Walter Schottky Institut and TUM School of Natural Sciences, Technische Universit\"at M\"unchen, Am Coulombwall 4, 85748 Garching, Germany.}
\author{Y. Tong}
\affiliation{Walter Schottky Institut and TUM School of Natural Sciences, Technische Universit\"at M\"unchen, Am Coulombwall 4, 85748 Garching, Germany.}
\author{A. Astaburuaga Hernandez}
\affiliation{Walter Schottky Institut and TUM School of Natural Sciences, Technische Universit\"at M\"unchen, Am Coulombwall 4, 85748 Garching, Germany.}
\author{P. Ji}
\affiliation{Walter Schottky Institut and TUM School of Natural Sciences, Technische Universit\"at M\"unchen, Am Coulombwall 4, 85748 Garching, Germany.}
\author{K. Gallo}
\affiliation{Department of Applied Physics, KTH Royal Institute of Technology, Roslagstullsbacken 21, Stockholm SE-106 91, Sweden.}
\author{A. V. Stier}
\affiliation{Walter Schottky Institut and TUM School of Natural Sciences, Technische Universit\"at M\"unchen, Am Coulombwall 4, 85748 Garching, Germany.}
\author{J. J. Finley}
\affiliation{Walter Schottky Institut and TUM School of Natural Sciences, Technische Universit\"at M\"unchen, Am Coulombwall 4, 85748 Garching, Germany.}
%
%##############################################################################
%
\date{\today}
%
%##############################################################################
%									ABSTRACT
%##############################################################################
%

\begin{abstract}
%Reducing dimensionality holds an excellent potential for creating and probing many-body quantum states with strong particle-particle interactions. 
We investigate the confinement of neutral excitons in a one-dimensional (1D) potential, engineered by proximizing hBN-encapsulated monolayer MoSe$_2$ to  ferroelectric domain walls (DW) in periodically poled LiNbO$_3$. Our device exploits the nanometer scale in-plane electric field gradient at the DW to induce the dipolar exciton confinement via the Stark effect. Spatially resolved photoluminescence (PL) spectroscopy reveals the emergence of narrow emission lines redshifted from the MoSe$_2$ neutral exciton by up to $\sim$100\,meV, depending on the sample structure. The spatial distribution, excitation energy response and polarization properties of the emission is consistent with signatures of 1D-confined excitons. The large electric field gradients accessible via proximal ferroelectric systems open up new avenues for the creation of robust quantum-confined excitons in atomically thin materials and their heterostructures.
\end{abstract}

%key words: Moir\'e lattice, Interlayer exciton emission, exciton-phonon interaction, phonon cascade .
%
%##############################################################################
%
\maketitle
%
%###############################################################################
%								MAIN TEXT
%###############################################################################
%

%\section{Introduction}
%\label{intro}

The confinement of particles to length scales comparable to their de Broglie wavelength leads to the quantization of their motional ground states~\cite{Bastard1992}. %serwane2011deterministic, kaufman2012cooling, gross2017quantum, bruzewicz2019trapped, kjaergaard2020superconducting, thureja2022electrically, heithoff2024valley}. 
When the thermal energy of the system, $E_T\simeq k_BT$, falls below the energy separation between these confinement-induced states, its quantum nature emerges, altering properties such as the energy spectrum of fundamental excitations. Reliable quantum technologies hinge on the precise manipulation of particles in this quantum regime~\cite{baugher2014optoelectronic, palacios2017large, branny2017deterministic, baek2020highly}. Furthermore, particle-particle interactions are significantly enhanced by reducing the dimensionality of a system~\cite{fogler2014high, unuchek2018room, seyler2019signatures, Shimazaki2019, wang2020correlated, xu2020correlated, tang2020simulation, regan2020mott, huang2021correlated, Li2021,  Wang2022, Campbell2022}, enabling the exploration of emergent quantum phases driven by the new interactions between these particles~\cite{carusotto2009fermionized, hallwood2010robust, carusotto2013quantum, noh2016quantum, schloss2016non, kennes2021moire, oldziejewski2022excitonic}. In this context, monolayer transition metal dichalcogenides (1L-TMDs) have emerged as a promising semiconductor platform due to their inherent two-dimensional (2D) confinement. These materials are direct band-gap at the $K/K'$ points of their hexagonal Brillouin zone~\cite{mak2010atomically, splendiani2010emerging,xiao2012coupled}, where interband optical transitions form tightly bound excitons~\cite{mak2013tightly, stier2016exciton, wang2018colloquium, stier2018magnetooptics, goryca2019revealing}. While excitons in 1L-TMDs couple strongly to light, achieving their motional quantization remains particularly challenging due to the heavy exciton masses and small exciton Bohr radius~\cite{stier2018magnetooptics,goryca2019revealing}. For example, an energy splitting between discrete confined motional ground states of $\hbar \omega \gtrsim 1\,$meV, requires a confinement length scale of $\ell_n=\sqrt{\frac{\hbar}{m_X\omega}} \lesssim 10\,$nm for an exciton mass ($m_X$) of the order of the free electron mass ($m_e$)~\cite{thureja2022electrically}.%=m_e+m_h$) of the order of the free electron ($m_e$)~\cite{thureja2022electrically}. 

The manipulation of excitons in 2D-semiconductor materials has largely centered around approaches such as moiré potential engineering~\cite{seyler2019signatures, zhang2021van, kennes2021moire, susarla2022hyperspectral, qian2024lasing}, nanopillars and strain engineering~\cite{palacios2017large, rosenberger2019quantum, kremser2020discrete, bai2020excitons, yu2021site, kogl2023moire}, electron and ion beam irradiation~\cite{fournier2021position, lenferink2022tunable}, and the local tuning of the dielectric environment~\cite{ugeda2014giant, stier2016exciton, raja2017coulomb, steinhoff2017exciton, forsythe2018band, price2019engineering,  moser2024atomically} to create interlayer junctions and trapping potentials. The use of partially overlapping gates~\cite{thureja2022electrically, heithoff2024valley, hu2024quantum}, has recently been shown to be a powerful approach to generate 1D exciton states~\cite{thureja2022electrically, heithoff2024valley} and to enable control of exciton wave functions~\cite{hu2024quantum}. This method utilizes the DC Stark shift induced by an in-plane electric field~\cite{cavalcante2018stark}, in combination with the formation of lateral $p-i-n$ junctions where the $p-$ and $n-$regions are defined by the gate arrangement~\cite{efimkin2017many, thureja2022electrically, heithoff2024valley, hu2024quantum} to create a total exciton confinement potential. This technique offers the advantage of in-situ tuning of the confinement potential from a continuum of 2D exciton states to the 1D quantum regime, where excitons are confined within potential traps of the order of $\sim$5\,meV. However, as the total confinement potential arises from the DC Stark shift and a sizeable repulsive Coulomb interaction~\cite{thureja2022electrically}, the motional ground states are susceptible to electronic noise and the composition of the 1D exciton wavefunction continuously varies with electron density and subband number \cite{thureja2022electrically}.

\begin{figure*}[ht!!]
\includegraphics*[keepaspectratio=true, clip=true, angle=0, width=2\columnwidth, trim={0mm, 0mm, 0, 0mm}]{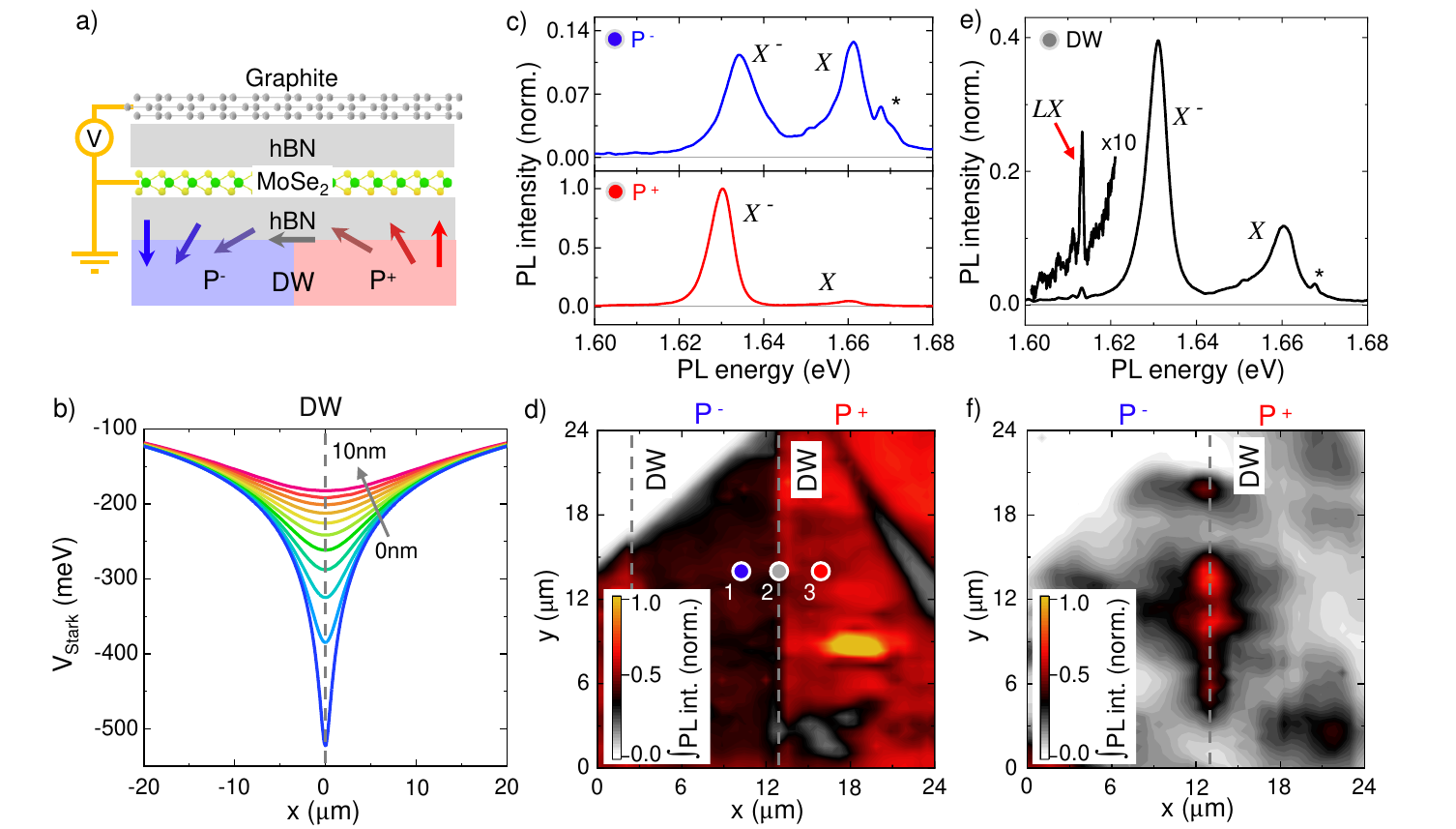}
\caption{\textbf{Sample description and emission signatures of 1D confineed excitons. a)} Schematic of the hBN-encapsulated 1L-MoSe$_2$ straddling DW boundaries in periodically poled LiNbO$_3$. The red-to-blue arrows indicate the direction of the PPLN electric field near the DW, which generates a 1D trapping potential ($V_{Stark}$). \textbf{b)} Calculated 1D trapping potential in the plane of the 1L-MoSe$_2$ for a variety of bottom hBN thicknesses. \textbf{c)} $\mu$PL spectra recorded at the $P^-$ and $P^+$ domains (top and bottom panels, respectively) showing the $X$ and $X^-$ emission. Features marked with an asterisk (*) correspond to Raman lines from the substrate. \textbf{d)} False colour map showing the integrated PL across the sample. The spatial modulation reveals the underlying PPLN domains. Spots 1 and 3 mark the sites in which spectra in d were acquired. \textbf{e)} $\mu$PL spectra recorded at the DW (spot 2 in d). In addition to the $X$ and $X^-$ emission, the spectra feature narrow emission lines ($LX$s) redshifted by $\sim50$\,meV from $X$. \textbf{f)} False colour map showing the background subtracted and integrated PL intensity of the $LX$.}
\label{figura:1}
\end{figure*}
%==FIGURA======FIGURA======FIGURA======FIGURA======FIGURA======FIGURA=====

 %\textbf{f)} False colour plot of the $LX$ spectral emission along the DW.

In this work, we demonstrate how nanometer scale ferroelectric domain boundaries in lithium niobate (LiNbO$_3$)~\cite{guarino2007electro, wang2018integrated, he2019high} can be used to induce robust 1D confinement of neutral excitons in 1L-MoSe$_2$, dominated by the Stark effect. LiNbO$_3$ is a very versatile ferroelectric material that can be integrated on oxide sacrificial layers using CMOS compatible processes to produce low loss waveguides~\cite{zhang2017monolithic} making it highly suitable for integrated optoelectronics and photonics~\cite{wen2019ferroelectric, li2016spatial, li2020polar, gallo2006bidimensional}. Furthermore, periodically poled lithium niobate (PPLN) exhibits large surface charge densities within individual ferroelectric domains and atomically sharp Neél-type domain walls (DW) between domains~\cite{cherifi2017non, jungk2014comment, morozovska2008effect}. Previously~\cite{soubelet2021charged}, we demonstrated that the in-plane electric field ($E_x$) at the DWs establishes a potential landscape like a lateral $p-n$ junction in 1L-TMDs when deposited on top of such domains. This potential induces 2D exciton drift and dissociation in the vicinity of the DW. The magnitude of $E_x$ was shown to be up to $E_x\sim400\,$V/$\mu$m and localized to lengthscales of few nanometers, far beyond what can be achieved with metallic contacts. This suggests that DWs can serve as quantum traps for excitons via the DC Stark effect~\cite{cavalcante2018stark}. To validate this hypothesis, we performed spatially resolved micro-photoluminescence ($\mu$PL) experiments, PL-excitation and polarization resolved spectroscopy on hBN encapsulated 1L-MoSe$_2$ samples transferred on top of a PPLN substrate. Our results reveal the formation of 1D exciton states localized at the DW, with narrow emission lines that are redshifted by up to $\sim$100\,meV from the MoSe$_2$ neutral exciton, depending on the sample structure. These observations are consistent with an exciton center-of-mass (COM) confined to lengthscales as small as $\sim$3\,nm. The observed absence of interactions with surrounding free charges, combined with the thermal robustness observed through temperature-dependent PL experiments, suggests that proximal electric fields arising from ferroelectric DWs may be  highly interesting for the exploration of strongly correlated exciton states~\cite{oldziejewski2022excitonic}.

\section{results and discussion}
\label{results}

Figure \ref{figura:1}a shows a schematic of a van der Waals layered device, representative of those used in this study. The detailed structure of each device is provided in the Supplemental Material (SM) 1~\ref{SN1}. The devices were assembled via dry viscoelastic stamping techniques~\cite{castellanos2014deterministic} and the design exploits the electric field at ferroelectric Néel-type DWs in LiNbO$_3$~\cite{cherifi2017non, liu2014theoretical}, depicted in Fig\ref{figura:1}a by the rotating arrows. The large and nanometer scale in-plane electric field ($E_x$) polarizes the TMD neutral exciton, resulting in a local reduction of the exciton transition energy and providing an attractive potential $V_{Stark}$ within the 1L-MoSe$_2$ directly above the DW via the DC Stark effect~\cite{cavalcante2018stark},
\begin{equation}
\label{eq:dcStark}
    V_{Stark} = -\frac{1}{2} \alpha {E_x}^2,
\end{equation}
where $\alpha=6.5$\,nm$^2$V$^{-2}$~\cite{cavalcante2018stark} is the in-plane exciton polarizability of MoSe$_2$. To estimate the in-plane electric field lengthscale and confinement potential, we calculated the electric field in the 1L-MoSe$_2$ plane as a function of the bottom hBN thickness using finite element simulations. The resulting potentials are shown in Fig.\ref{figura:1}b for a bottom hBN thickness varying from 0\,nm to 10\,nm and highlights the possibility to tune $V_{Stark}$ by selecting the bottom hBN thickness.
%and solve the eigenenergy equation for the first 5 confined states ($\psi_0$, $\psi_1$, $\psi_2$) in the Supplemental Material Fig. S7.
Further details on the simulations of the PPLN electric field and the trapping potential are provided in Section \ref{Methods} and the SM 1~\ref{SN1}. To tune the electronic landscape in the TMD, the 1L-MoSe$_2$ was encapsulated in thin flakes of hexagonal boron nitride (hBN) and a top gate was incorporated to the structure by using a thin graphite flake. The Ohmic contact to the 1L-MoSe$_2$ was oriented across the DW to ensure that all regions of the sample are grounded, regardless of the junction potential.
% completing the device architecture. 
%In a previous study~\cite{soubelet2021charged}, we demonstrated that the combined structure of the DW and MoSe$_2$ behaves like a nanometer scale p-n homojunction. Therefore, this configuration ensures that all regions of the sample are grounded, regardless the junction potential. 

%For the interested readers, Although all the experiments discussed in this letter were conducted on the same sample with an 8\,nm thick hBN bottom layer, the main experimental observations were consistently reproduced across multiple samples. Figure \ref{figura:1}b shows an optical micrograph of that sample, with the green lines marking the contour of the 1L-MoSe$_2$. The yellow horizontal stripes correspond to gold pads used to contact the graphite layers. The PPLN DWs, though not optically visible, are oriented perpendicular to the gold pads, completing the device architecture. In a previous study~\cite{soubelet2021charged}, we demonstrated that the combined structure of the DW and MoSe$_2$ behaves like a nanometer scale p-n homojunction. Therefore, this configuration ensures that all regions of the sample are grounded, regardless the junction potential. 

%==FIGURA======FIGURA======FIGURA======FIGURA======FIGURA======FIGURA=====
\begin{figure*}[t!!]
\includegraphics*[keepaspectratio=true, clip=true, angle=0, width=2\columnwidth, trim={0mm, 0mm, 0mm, 0mm}]{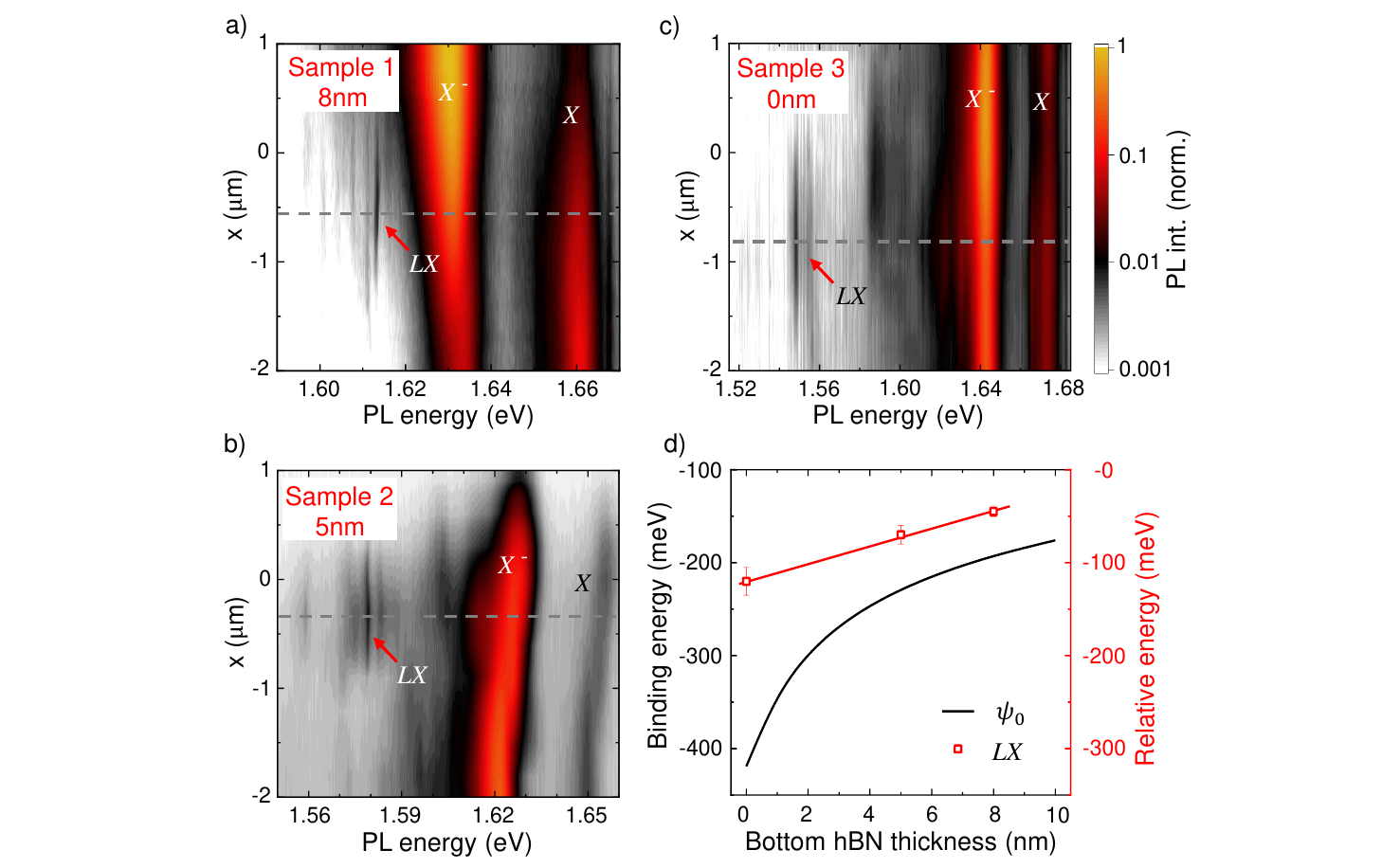}
\caption{\textbf{PL experiments across samples with different bottom hBN thickness} \textbf{a)}, \textbf{b)} and \textbf{c)} False-color plot of the 1L-MoSe$_2$ emission across the DW for three different samples. Sample 1 is a 1L-MoSe$_2$ stacked on top of 8\,nm thick hBN, Sample 2 a 1L-MoSe$_2$ stacked on top of 5\,nm thick hBN, and sample 3 is a 1L-MoSe$_2$ directly stacked on top of the PPLN. The grey dotted line marks the DW position \textbf{d)} Comparison between the theoretical binding energy of the first confined state ($\psi_0$) and the experimentally observed localized states ($LX$) energy relative to the $X$ energy for the three samples. The red line over the experimental data serves as a guide to the eye.
}
\label{figura:2}
\end{figure*}
%==FIGURA======FIGURA======FIGURA======FIGURA======FIGURA======FIGURA=====

We begin our investigation with spatially resolved low-temperature ($T=4.2\,$K) PL spectroscopy on a sample with 8\,nm thick bottom hBN (Sample 1). The sample was excited using a CW optical power $P_{ex} = 1\,\mu$W, focused to a diffraction-limited spot of $\sim500\,$nm (100$\times$ objective, NA = 0.82). The excitation photon energy was set to $E_{ex}=1.722$\,eV, $\sim$60\,meV above 1L-MoSe$_2$ neutral exciton ($X$). Figure \ref{figura:1}c shows selected PL spectra corresponding to two distinct domains with opposite out of plane ferroelectric polarization ($P^-$ and $P^+$, respectively)\cite{soubelet2021charged}. The main difference between these spectra originates in the modulation of the free charge density across the domains. From the ratio of the trion ($I_{X^-}$) to exciton ($I_X$) intensities, we estimate a negative free charge density of $n_{P^-} \simeq 0.9 \times 10^{10}$\,cm$^{-2}$ and $n_{P^+} \simeq 3.6 \times 10^{10}$\,cm$^{-2}$ for the $P^-$ and $P^+$ domains, respectively~\cite{ross2013electrical}. Figure \ref{figura:1}d shows a false-colour map of the integrated intensity of the 1L-MoSe$_2$ emission across the device. The spectra in Fig.~\ref{figura:1}c were obtained from the locations marked as ``1'' (blue dot) and ``3'' (red dot) in Fig.~\ref{figura:1}d. The PL intensity directly reflects the underlying PPLN structure, making the PL map an effective tool for identifying the ferroelectric domains. Based on the spatial modulation of the PL intensity, the ferroelectric polarization of each domain was determined in accordance with our previous work~\cite{soubelet2021charged}. 

The key observation is shown in figure \ref{figura:1}e where we plot the spectrum recorded directly at the DW (grey spot ``2'' in Fig.~\ref{figura:1}d) and reveals narrow emission lines ($LX$s) redshifted by $\sim$50\,meV relative to $X$. We follow the spatial distribution of these emission features by integrating the 1L-MoSe$_2$ emission over the $LX$ spectral range (1.595\,eV to 1.615\,eV), while subtracting the broad $X^-$ emission as a background. The resulting map is shown in figure \ref{figura:1}f, indicating that these spectrally narrow emission lines are macroscopically localized along the DW. For the remainder of this manuscript, we demonstrate that the $LX$ emission is consistent with the radiative recombination of excitons confined in a 1D quantum trap at the DW.

The $LX$-emission lines were observed in a variety of samples with slightly different layer structure and Ohmic contacts. Figure \ref{figura:2}a to c summarizes low-T $\mu$-PL linescans across the DWs in three different samples (details in SM 2~\ref{SN2}). In each panel, the location of the DW is indicated with a grey dashed line. The main difference between the samples is the bottom hBN thickness, as indicated in the panels. The PL of all samples show the neutral exciton emission $X$, the trion ($X^-$) and, diffraction limited at the DW, the $LX$ emission with an energy relative to $X$ of $\sim-50\,$meV, $\sim-70\,$meV and $\sim-120\,$meV, for samples 1 to 3, respectively. The PL experiments performed on these samples show that by reducing the bottom hBN thickness, the emission energy of the $LX$ redshifts with respect to $X$, directly showing the tunability of the confinement potential by controlling $E_x$ in the plane of the 1L-MoSe$_2$. To validate this result, we numerically solved the Schr\"odinger equation for the potential traps shown in Fig.~\ref{figura:1}b and obtained the wavefunctions and eigenenergies of the system (See SM3~\ref{SN3} for details). Figure \ref{figura:2}d shows the calculated binding energy of the confined ground state in the quantum trap ($\psi_0$) and the relative energy of the $LX$s with respect to the observed $X$ energy. The overall trend follows the simulations, although the model overestimates the precise $LX$ confinement. Specifically, the calculated binding energy is $\sim300$\,meV higher for thin hBN thicknesses and $\sim100$\,meV higher for hBN thicknesses $>5$\,nm. This mismatch is possibly due to our oversimplified dielectric model, which is not valid for very thin hBN layers~\cite{slobodeniuk2023exciton} and other systematic surface charge potentials, the detailed origin of which are beyond the scope of this work.

\begin{figure*}[t!!]
\includegraphics*[keepaspectratio=true, clip=true, angle=0, width=2.\columnwidth, trim={0mm, 0mm, 0mm, 0mm}]{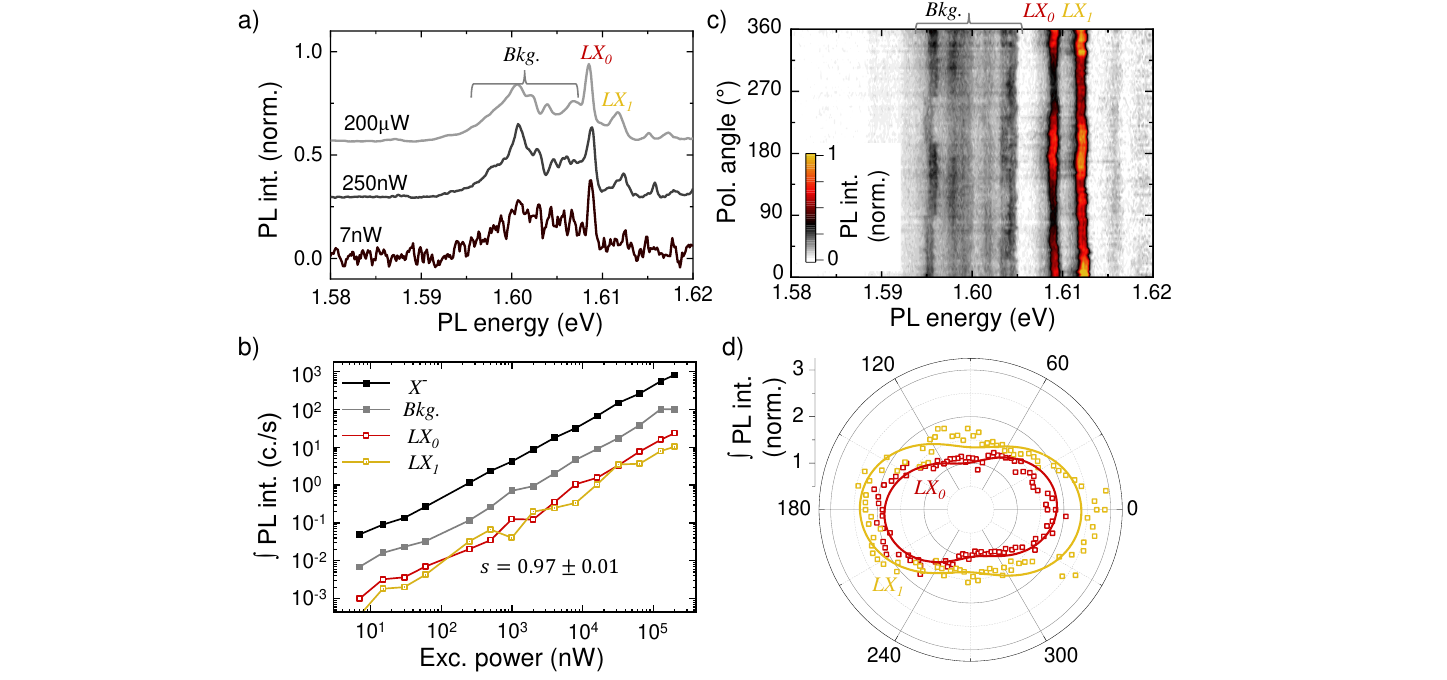}
\caption{\textbf{Power-dependency and linearly polarized emission of 1D confined states. a)} Tron-subtracted $LX$ PL spectra in resonance under varying excitation powers, 7\,nW, 250\,nW, and 200\,$\mu$W. \textbf{b)} Integrated intensity of $X^-$, Background (Bck.), and individual $LX$s as a function of excitation power. The intensity follows a power-law dependence with an exponent $s=0.97\pm0.05$. \textbf{c)} False-color plot of the background-subtracted in resonance $LX$ PL spectra as a function of the excitation-collection linear polarization angle. \textbf{d)} Polar plot of the PL spectral intensity at the central position of individual $LX$s as a function of excitation-collection linear polarization angle. Solid lines are a sinusoidal fit.}
\label{figura:3}
\end{figure*}
%==FIGURA======FIGURA======FIGURA======FIGURA======FIGURA======FIGURA=====

%Although the experiments discussed hereafter in this letter were conducted on the same sample with an 8\,nm thick hBN bottom layer
We continue the discussion of the 1D-nature of the $LX$ states through the data obtained from sample 1 (8\,nm bottom hBN), although similar results have been obtain from the other samples. In MoSe$_2$, localized emission from defects occur within a similar energy range than the observed $LX$ emission~\cite{chakraborty2016localized}. Therefore, we conducted power-dependent PL experiments to distinguish between those possibilities. 
%As we discuss later in the text, the intensity of the $LX$s emission exhibits a pronounced resonance when excited close to the $X$ energy, consequently these experiments were conducted exciting at $E_{ex}=1.66$\,eV. 
Figure \ref{figura:3}a shows selected PL spectra of the $LX$ emission, by resonantly exciting $X$ at $E_{ex}=1.66$\,eV, with varying excitation power $P_{ex}$ from 7\,nW to 200\,$\mu$W. At very low $P_{ex}$, the spectrum is comprised of a single narrow emission line (linewidth $\sim400\,\mu$eV, $LX_0$) superimposed on a broad background. With increasing $P_{ex}$, the background develops some structure while maintaining its relative intensity with respect to the $LX_0$ line. In addition, a few narrow peaks, spaced by $\sim3$\,meV emerge from the noise floor that are blue detuned from $LX_0$ (e.g. $LX_1$). Figure \ref{figura:3}b shows the integrated PL intensity of the distinct spectral features as a function of $P_{ex}$. Notably, the background, $LX_0$, and $LX_1$ exhibit the same power-law dependence as the free trion, scaling approximately linear (s=$0.97\pm0.01$). The absence of any saturation over more than five orders of magnitude of $P_{ex}$ indicates that the $LX$ emission is not defect-like~\cite{chakraborty2016localized}.% While the overall power dependence of $LX_0$ and $LX_1$ follows a linear trend with $P_{ex}$, their relative intensity depends on $P_ex$. \textcolor{red}{and that measn what?}

%oscillates, and depending on the power range, $LX_1$ dominates over $LX_0$. %This observation supports the hypothesis of two electronic levels competing for the same exciton population. Further details can be found in SM XX \ref{}.

%\hl{This observation supports the hypothesis of interacting particles confined within the same potential trap that exhibit an ordering influenced by their 1D density.}exhibit circular optical selection rules

Free excitons in TMDs exhibit circular optical selection rules ~\cite{wang2018colloquium, cao2012valley, mak2012control}, however, excitons confined within a 1D channel display a distinct behaviour. As previously reported, the long-range electron-hole exchange interaction mixes the exciton center-of-mass (COM) motion with the $K/K'$-valley degrees of freedom~\cite{thureja2022electrically, heithoff2024valley, wang2021highly}. The motional in-plane anisotropy from the 1D trapping potential leads to the splitting of the confined exciton COM wavefunctions into orthogonal linearly polarized states, parallel or perpendicular to the DW. Consequently, the energy of the $n^{th}$ quantized state splits into two components, one with linear polarization parallel to the DW and energy $E_n$, and a second linearly polarized perpendicular to the DW and energy $E_n^{\perp} = E_n + \delta_n$. Here, $\delta_n = \gamma k_n$, where $\gamma$ is the exchange-interaction coupling parameter and $k_n$ the average COM wavevector of the state~\cite{heithoff2024valley, glazov2015spin, wang2021highly}.

Motivated by this expectation, we performed polarization-resolved colinear excitation-collection PL spectroscopy on the localized exciton peaks at the DW. Figure \ref{figura:3}c shows a false-color plot of the background subtracted $LX$s emission as a function of the excitation-emission angle measured relative to the direction parallel to the DW. The $LX_0$ and $LX_1$ states exhibit a clear linear polarization parallel to the DW, with significantly reduced emission perpendicular to the DW. Figure \ref{figura:3}d shows the extracted intensity of $LX_0$ and $LX_1$ as a function of the excitation-collection angle, solid lines correspond to sinusoidal fits to the data. Both $LX_0$ and $LX_1$ display intensity suppression in the direction perpendicular to the DW of $\sim50\%$ and $\sim40\%$, respectively. This behaviour is consistent with previous observations of 1D confined excitons in TMDs~\cite{thureja2022electrically, heithoff2024valley}, although the degree of suppression observed here is comparatively lower. The absence of emission in the direction perpendicular to the DW aligns with previous reports of 1D states. As the confinement increases, the corresponding $k_n$ increases, shifting states with perpendicular polarization to higher energies, thereby rendering their occupation unfavourable and eventually suppressing their emission~\cite{thureja2022electrically}. In contrast, the background emission shows a significantly smaller modulation in its intensity and, in the case of the lower energy emission component, a spectral shift (see Fig.~\ref{figura:3}c). The polarization-resolved behaviour of the background shows a clearly different behavior as the $LX$. While a precise identification of these background states remains unclear and demands further understanding, they may stem from strain, local dielectric fluctuations, or structural imperfections, which modulate the 1D-trapping potential along the DW.

%deviates from that expected for 1D exciton states, \textcolor{red}{I don't undrstand the statement here}reinforcing our hypothesis that it arises from variations of the 1D channel along the DW. The precise nature of these background states remains unclear and demands further understanding.

\begin{figure*}[t!!]
\includegraphics*[keepaspectratio=true, clip=true, angle=0, width=1\linewidth, trim={0mm, 0mm, 0mm, 0mm}]{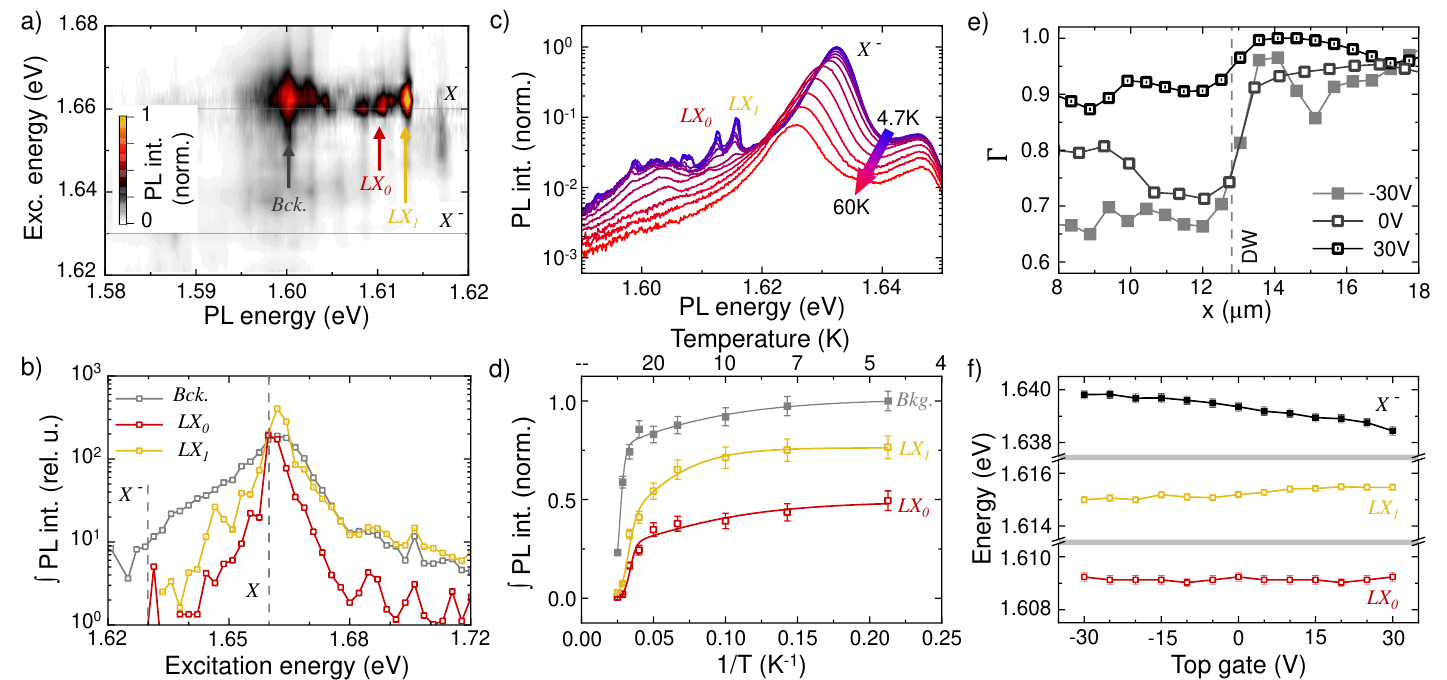}
\caption{\textbf{Resonance effects, temperature evolution and free charge density effects on the 1D confined states. a)} False-colour plot of the trion-subtracted $LX$ PL spectra as a function of the excitation energy revealing the resonance with $X$. \textbf{b)} Integrated PL spectral intensity as function of the excitation energy. \textbf{c)} In resonance $LX$s and trion PL spectra as a function of temperature, from 4.7\,K to 60\,K. \textbf{d)} Integrated PL intensity as a function of $1/T$. Solid lines correspond to a modified Arrhenius fitting. \textbf{e)} Line scan showing the top gate voltage effect on the 1L-MoSe$_2$ photophysics described through the parameter $\Gamma= I_{X^-}/( I_{X^-}+ I_{X})$. \textbf{f)} $X^-$, $LX_0$ and $LX_1$ spectral position as a function of the top gate voltage.}
\label{figura:4}
\end{figure*}

To explore the 1D-exction levels and their relationship with the 2D counterpart, we conducted PL-excitation (PLE) experiments. Figure \ref{figura:4}a shows a false-colour plot of the $LX$ PL spectra as a function of $E_{ex}$. All features in this spectral range exhibit a strong resonance when exciting near the $X$ energy at $\sim 1.66\,$eV, indicated by the grey horizontal line. No resonance is observed when the excitation laser energy is swept across the $X^-$. Figure \ref{figura:4}b shows the integrated PL intensity of the spectral lines (arrows in Fig.\ref{figura:4}a) as a function of excitation energy. Resonant excitation of the $X$ increases the $LX$s emission by two orders of magnitude as compared to non-resonant excitation. This shows that the most efficient way to optically excite the 1D channel is by generating a 2D exciton population which is subsequently trapped at the DW. Notably, even while exciting the sample at $E_{ex} = 1.62$\,eV, $\sim10\,$meV above $LX_0$, there is no increase in the emission intensity of any features. This observation suggests two possible explanations, the relatively small area of the 1D channel compared to the overall laser beam size, and the inherently lower oscillator strength of polarized excitons relative to $X$.
%\textcolor{red}{this text has to be rewritten} On the one side, the relatively small illuminated area of the DW region, compared to the overall beam size. On the other hand, the $LX$, being polarized excitons, would be expected to exhibit a lower oscillator strength compared to $X$s.

To investigate the thermal robustness of the 1D confined states, we performed temperature dependent PL. Figure \ref{figura:4}c shows temperature-dependent PL spectra, from $T=4.7\,$K to $60\,$K, while exciting the sample resonantly with the $X$ energy. The $X^-$ emission remains unaffected in both intensity and spectral position until the temperature reaches approximately 40\,K. Beyond this point, increasing temperature causes the $X^-$ to notably redshift and decrease in intensity, consistent with previous reports~\cite{jadczak2017probing}. In contrast, the intensity of the $LX$s and the background gradually decrease, with a sudden variation above 30\,K. Figure \ref{figura:4}d presents the extracted intensities of $LX_0$, $LX_1$, and the background as a functions of $1/T$. The temperature dependence of each feature [$I(T)$] is fitted with a modified Arrhenius equation that incorporates two activation energies, $E_1$ and $E_2$:
\begin{equation} 
\label{eq:Arrhenious}
     I(T) = \frac{I_0}{1+A e^{\left(-E_1/k_BT\right)} + B e^{\left(-E_2/k_BT\right)}},
\end{equation}
where $A$ and $B$ are amplitude parameters. The fitting procedure yields a first activation energy of $E_1\sim2\,$meV for all features, which governs the intensity evolution at low temperatures. The second activation energy is determined to be $47\pm4\,$meV, $40\pm5\,$meV, and $32\pm5\,$meV for the background, $LX_0$, and $LX_1$ features, respectively. Notably, in all cases, the second activation energy closely matches the energy difference to the $X$ and not the nearby $X^-$, indicating the activation of a scattering mechanism that facilitates the transition of these localized states into the 2D continuum.

%==FIGURA======FIGURA======FIGURA======FIGURA======FIGURA======FIGURA=====
%==FIGURA======FIGURA======FIGURA======FIGURA======FIGURA======FIGURA=====

Lastly, we analyzed the influence of free charges on the $LX$ features by tuning the electron density in the sample via the top gate (see Fig.~\ref{figura:1}a). The effect of the charge density is characterized through the parameter $\Gamma = I_{X^-}/(I_{X^-}+I_X)$, where $I_{X^-}$($I_{X}$) is the trion(exciton) intensity. Therefore, in a charge neutrality condition, $\Gamma\to 0$ since $I_{X^-}\to 0$. On the other hand, by increasing the free charge density, $\Gamma\to 1$ since $I_X\to 0$. The effect of the top gate is summarized by the line scans in Figure \ref{figura:4}e, which show $\Gamma$ in the vicinity of the DW for three selected voltages (for further details, see SM \ref{SN4}). While the 1L-MoSe$_2$ is uniformly gate-biased, the effect of the charging is notably portrayed by the behavior of $\Gamma$ on the $P^-$ domain. Figure \ref{figura:4}f shows the extracted spectral position for $LX_0$, $LX_1$ and $X^-$ as a function of the top gate voltage at the DW. The trion feature redshifts by $\sim$2\,meV throughout the top gate voltage range, consistent with the behavior of repulsive polarons with increasing the 2D Fermi level~\cite{ross2013electrical}. Meanwhile, the $LX$s spectral position remains nearly constant, indicating that the trapping potential is unaffected by the free charge density. This behaviour contrasts the work on gate-induced exciton confinement \cite{thureja2022electrically, heithoff2024valley} where the effective trapping potential is a combination of Stark effect and interaction-induced confinement, which depends on the free carrier density. Therefore, the confinement potential in our platform is not a result of many-body interactions and is, consequently, insensitive to variations in the local free charge density. %\textcolor{red}{if we are over length, we can cut the next sentence}This resilience highlights potential optoelectronic applications where charge density governs the photophysics of the 2D material while leaving the 1D states unaltered.

We conclude by discussing the length scales of the confinement potential. Our calculations (see Fig.~\ref{figura:1}b) show that the maximum confinement energy for an encapsulated 1L-MoSe$_2$ sample with an $8\,$nm thick bottom hBN is $-200\,$meV relative to $X$ (see SM 3~\ref{SN3} for details). However, as discussed above, $LX$s in this device are redshifted by only $-50\,$meV. This difference suggest that the built-in electric field generated by the $p-n$ homojunction at the DW diminishes $V_{Stark}$ and results in a shallower effective potential $V_{eff}$. Therefore, to calculate the confined COM wavefunctions, we rescale $V_{Stark}$ to a depth of 50\,meV while preserving its overall profile. By numerically solving the Schr\"odinger equation for $V_{eff}$ using a total exciton effective mass $m_{X} = 1.29m_0$, where $m_0$ is the electron mass~\cite{LI20231312}, we obtained the wavefunctions and eigenenergies of the system (see SN 3~\ref{SN3} for details). The COM confinement, defined as $\ell_n=\sqrt{<\psi_n|x^2|\psi_n>}$, yields $\ell_0 = 3.0\,$nm and $\ell_1 = 5.6\,$nm for the ground and first excited confined states, with an energy separation of $\Delta E = 4.5$\,meV. Assuming the experimentally observed spectral features $LX_0$ and $LX_1$ corresponds to these states, the measured value of $\Delta E \sim 3.5$\,meV, which slightly depends on the specific position along the DW, is in good agreement with theoretical expectations. Using this experimental $\Delta E = \hbar\omega$ within an harmonic confinement potential approximation, we estimate the COM confinement lengthscale to be $\ell_n=\sqrt{\frac{\hbar}{m_X\omega}\left(n+\frac{1}{2}\right)}$, yielding $\ell_0 = 2.9\,$nm and $\ell_1 = 5.0\,$nm. Notably, both the harmonic approximation and the numerical estimation of $V_{eff}$ yield similar COM confinement values. More important, our device structure effectively decouples the 1D exciton state from the 2D counterpart by providing a potential trap that, compared to prior studies~\cite{heithoff2024valley, thureja2022electrically}, is an order of magnitude higher with a $\sim50\%$ smaller COM confinement length.

%We explain the experimentally observed slight variations of the energy splitting along the DW with local potential variations due to the local dielectric environment, adhesion of the stack to the substrate etc. (see, e.g., Fig.~\ref{figura:3}c, \ref{figura:4}a and \ref{figura:4}c).
%Taking the in-plane exciton polarizability in MoSe$_2$ to be $\alpha=6.5$nm$^2$V$^{-2}$~\cite{cavalcante2018stark} and the derived value of $E_x(x)$, we calculate the dipolar trapping potential $V_{Stark}(x)$ (see \ref{figura:1}b).

\section{Conclusions}
\label{Conclusions}

In summary, we fabricated optoelectronic devices that leverage the large in-plain electric field gradients at the PPLN DWs to confine 1L-MoSe$_2$ neutral excitons within a 1D channel. Spatially resolved $\mu$-PL experiments revealed narrow emission lines at the DW, redshifted from the neutral 1L-MoSe$_2$ exciton. These emission lines appear diffraction-limited in the direction perpendicular to the DW while extend macroscopically along the DW. Complementary power-dependent PL, linearly polarized PL, PLE, and temperature dependent PL spectroscopy suggest that these emission lines are consistent with the formation of a 1D exciton gas at the DW. Although our design currently lacks a mechanism to fine-tune the confinement potential in-situ, the proper selection of the bottom hBN layer thickness allows manipulation of the confinement potential by up to $\sim100$\,meV, as demonstrated in a sequence of samples. This is an order-of-magnitude enhancement as compared to previous reports~\cite{heithoff2024valley, thureja2022electrically}. This robust confinement effectively decouples the 1D exciton state from its 2D counterpart and suppresses many-body interactions with the environment outside the 1D channel. As a result, our platform offers a compelling system for manipulating and localizing excitons in TMDs, enabling the future exploration of 1D exciton dynamics as well as highly correlated phases of 1D-dipolar exciton gases~\cite{carusotto2009fermionized, hallwood2010robust, schloss2016non, oldziejewski2022excitonic}.

%Our results underscore the need for further optimization of the device structure to enable in situ tuning of the confinement potential depth and to fully exploit the ferroelectric grain boundaries. This approach provides a pathway toward the development of integrated photonic structures and waveguides combined with 2D materials. By strategically designing the ferroelectric domains, which can be freely tailored during the poling process~\cite{wen2019ferroelectric, li2016spatial, li2020polar, gallo2006bidimensional}, this platform offers the ability to arbitrarily shape the potential landscape for excitons, unlocking new opportunities for advanced excitonic devices and applications.

\section{Methods}\label{Methods}

\subsection{Sample Fabrication}

The periodically poled domains were fabricated by bulk electric field poling of commercial 500$\mu$m thick congruent $z$-cut LiNbO$_3$ crystals. The quality of the poling was
verified optically, which allows us to visualize the domain pattern and evaluate its uniformity without any surface contamination or disruption~\cite{missey2000real}. Furthermore, reference PPLN samples fabricated in the same batch were etched for 10\,min in HF acid to convert the PPLN pattern into a surface relief grating (due to the differential etching of $−z$ and $+z$ domains in HF), further confirming the successful domain inversion on both faces with a domain duty cycle close to 50\% on the lithographically patterned face of the sample (originally $−z$). 

Monolayers MoSe$_2$, hBN flakes and few layers graphene were obtained from commercial bulk crystals via mechanical exfoliation. We specifically selected 1L-MoSe$_2$ due to its lack of dark states below the $X$ energy and the simplicity of its PL spectra at low temperature, therefore facilitating the identification of additional emission features that result from the 1D $X$ confinement. 

All the samples used in this work were stacked and encapsulated between thin hBN flakes using dry transfer techniques based on polycarbonate films, similar to Ref. \onlinecite{castellanos2014deterministic}.

\subsection{Optical experiments}

All experiments were conducted using a helium exchange gas cryostat, equipped with a cryogenically compatible objective (numerical aperture $NA=0.82$) and a temperature controller. The excitation source was a CW tunable Ti:Sa laser.

%
%##############################################################################
%    Acknowledgements & Contributions
%##############################################################################
%

\section{Acknowledgements}

We gratefully acknowledge the German Science Foundation (DFG) for financial support via the SPP-2244 (DI 2013/5-1, FI 947/7-2, FI 947/7-1 and FA 971/8-1) the clusters of excellence MCQST (EXS-2111) and e-conversion (EXS-2089). P.S. acknowledges the financial support from the DFG through the Walter Benjamin program.

\section{Author contribution}

P.S., A.V.S., and J.J.F. conceived the project and K.G. provided the PPLN substrates. P.S. modeled the system and developed the calculations with the participation of A.A.H. P.S. designed the samples, which were fabricated by P.S., Y.T., and P.J. P.S., Y.T. and A.A.H. performed the optical measurements, and P.S. and Y.T. analyzed the data. P.S. and A.V.S. wrote the paper with input from all coauthors. All authors reviewed the manuscript.

\section{Data availability}

The data that support the findings of this study are available on request from the corresponding author.

\section{Competing interests}

The Authors declare no Competing Financial or Non-Financial Interests.

%\clearpage  

%
%##############################################################################
%    Additional information
%##############################################################################
%

\onecolumngrid

\section{Supplementary Material}

\subsection{Determination of the electric field at the domain wall}\label{SN1}

\begin{figure}[t!!]
\includegraphics*[keepaspectratio=true, clip=true, angle=0, width=1.\columnwidth, trim={1mm, 95mm, 38, 2mm}]{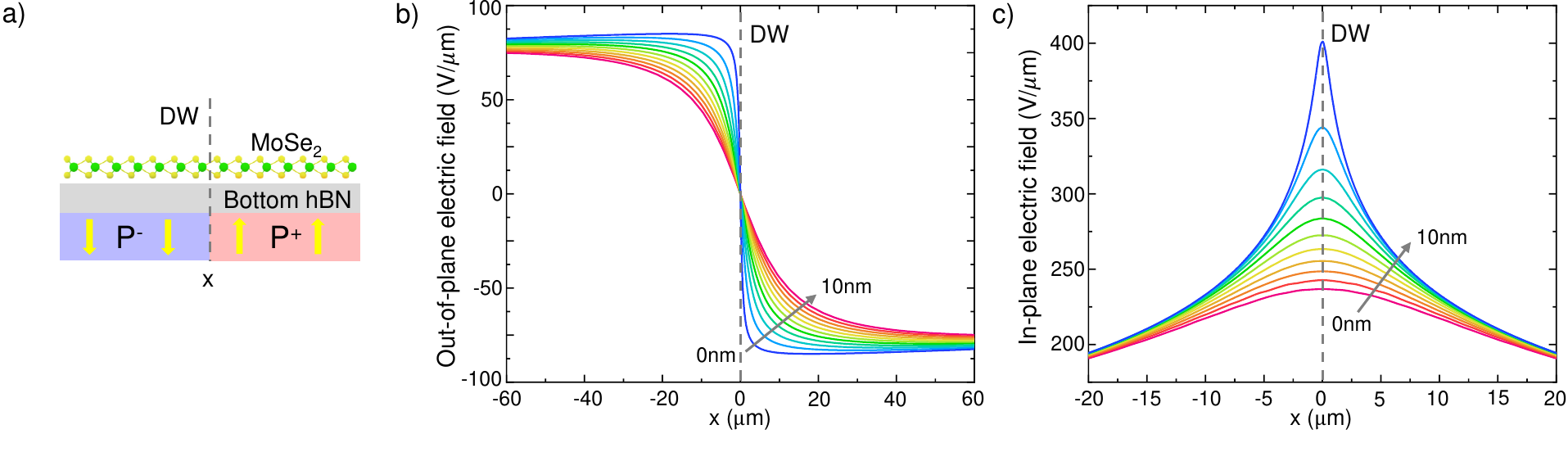}
\caption{\textbf{Calculated electric field at the DW as a function of the bottom hBN.} \textbf{a)} Schematic representation of the bottom hBN flake and MoSe$_2$ monolayer used to calculate the electric field at the PPLN DW. \textbf{b)} Calculated out-of plane electric field component. \textbf{c)} Calculated in-plane electric field component.}
\label{figura:SM1}
\end{figure}

Figure \ref{figura:SM1}a displays a schematic representation of the PPLN substrate, the bottom hBN flake and the monolayer 1L-MoSe$_2$ used to estimate the electric field at the TMD position. Figures \ref{figura:SM1}b and c present the out-of-plane and the in-plane electric field at the TMD monolayer position as a function of the bottom hBN used for the sample encapsulation. The calculations were produced by finite elements assuming an hBN refractive index of $n_{hBN}=1.8$ and an effective surface charge density on the PPLN domains of $\pm$2.7\,$\mu$C/$\mu$m$^2$ as determined in our previous work~\cite{soubelet2021charged}.

Note that, by grounding the samples, the reordering of charges screens the electric field along all domains, except the in-plane electric field at the DW, where the device behaves like a nanometer scale $p-n$ homojunction generating a build in electric field that reduces $V_{Stark}$ resulting in an effective potential $V_{eff}$.

%To avoid exciton dissociation, we estimated the minimum hBN thickness such that the in-plane electric field at the DW $E_x(0) . d_X . 2q_e < E_{bind}$, where $d_X\simeq1\,$ is the exciton size $q_e$ the electron charge and $E_{bind}\simeq 200\,$meV the exciton binding energy. 

%==FIGURA======FIGURA======FIGURA======FIGURA======FIGURA======FIGURA=====
%==FIGURA======FIGURA======FIGURA======FIGURA======FIGURA======FIGURA=====

\subsection{Observation of sharp emission lines associated to 1D confined excitons in different samples}\label{SN2}

\begin{figure}[t!!]
\includegraphics*[keepaspectratio=true, clip=true, angle=0, width=.85\columnwidth, trim={7mm, 21mm, 17, 0mm}]{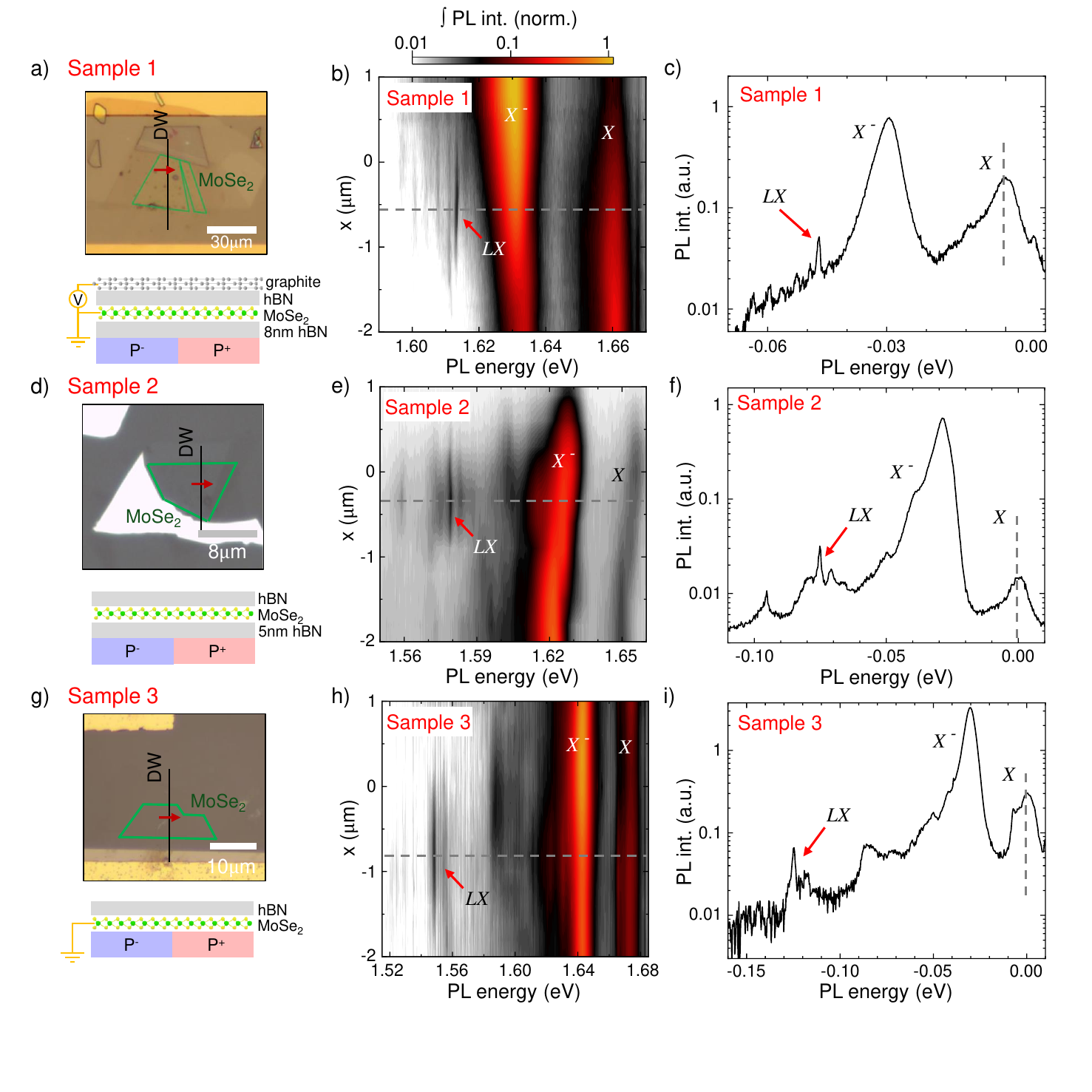}
\caption{\textbf{PL experiments on different samples.} \textbf{a)} Optical micrograph and schematic representation of the 1L-MoSe$_2$ sample presented in the main text and that was stacked on top of 5\,nm bottom hBN flake. \textbf{b)} False-colour plot of the integrated PL intensity across the DW for the sample in a. \textbf{c)} PL spectra at the DW for the sample in a. \textbf{d)} Optical micrograph of a 1L-MoSe$_2$ stacked on the PPLN with a 5\,nm bottom hBN flake. \textbf{e)} False-colour plot of the PL intensity across the DW for the sample in d. \textbf{f)} PL spectra at the DW for the sample presented in d. \textbf{g)} Optical micrograph of a grounded 1L-MoSe$_2$ stacked directly stacked on the PPLN. \textbf{h)} False-color plot presenting the PL intensity across the DW for the sample in g. \textbf{i)} PL spectra at the DW for the sample in g. The PL spectra in c, f and i are presented on an energy scale relative to the 2D neutral exciton.}
\label{figura:SM2}
\end{figure}

The observation of sharp emission lines attributed to localized exciton states was consistently reproduced across multiple samples. The results are summarized in Figure \ref{figura:SM2}, which presents data from three distinct samples. Specifically, Figs.~\ref{figura:SM2}a, d, and g displays the optical micrograph of each device, with the green lines marking the contour of the 1L-MoSe2 and a vertical black line marking the DW on each sample. Additionally, below each optical picture, a schematic representation shows the sample architecture. Figs. \ref{figura:SM2}b, e, and h presents the false-color plot of their PL (samples in Figs.~\ref{figura:SM2}b, e, and h, respectively) across the DW and along the red arrows depicted in their micrographs. Finally, Figs.~\ref{figura:SM2}c, f, and i present, for each device in Figs.~\ref{figura:SM2}a, d, and g, respectively, the PL spectra at the DW. The energy scale in these figures is relative to the 2D neutral exciton.

The 1L-MoSe$_2$ flake in the sample of Fig.~\ref{figura:SM2}a correspond to the grounded sample presented in the main text, it was stacked with 8\,nm bottom hBN, and its false color plot in Fig.~\ref{figura:SM2}b and its PL in c display $LX$s emission redshifted by $\sim$50\,meV from the exciton feature. The sample in Fig.~\ref{figura:SM2}d has 5\,nm bottom hBN and was not grounded. As a result, $X$ and $X^-$ spectrally shift due to the out-of-plane electric field that is not completely screened and produces Stark shift. Its false colour plot in Fig.~\ref{figura:SM2}e and its PL in f show the $LX$s redshifted by $\sim$70\,meV from the exciton feature. The last sample, displayed in Fig.~\ref{figura:SM2}g was directly stacked on top of the PPLN and was grounded. Its false colour plot in Fig.~\ref{figura:SM2}h and the spectra i display the exciton and trion features and, $\sim$120\,meV redshifted from the exciton, the localized states. 

Our results across different samples show that it is possible to tune the 1D confinement by properly selecting the bottom hBN thickness in the device structure (see next section for details). 

%==FIGURA======FIGURA======FIGURA======FIGURA======FIGURA======FIGURA=====
%==FIGURA======FIGURA======FIGURA======FIGURA======FIGURA======FIGURA=====

\subsection{Eigenfunctions and eigenenergies of the 1D device}\label{SN3}

\begin{figure}[t!!]
\includegraphics*[keepaspectratio=true, clip=true, angle=0, width=1\columnwidth, trim={-50mm, 95mm, 340, 0mm}]{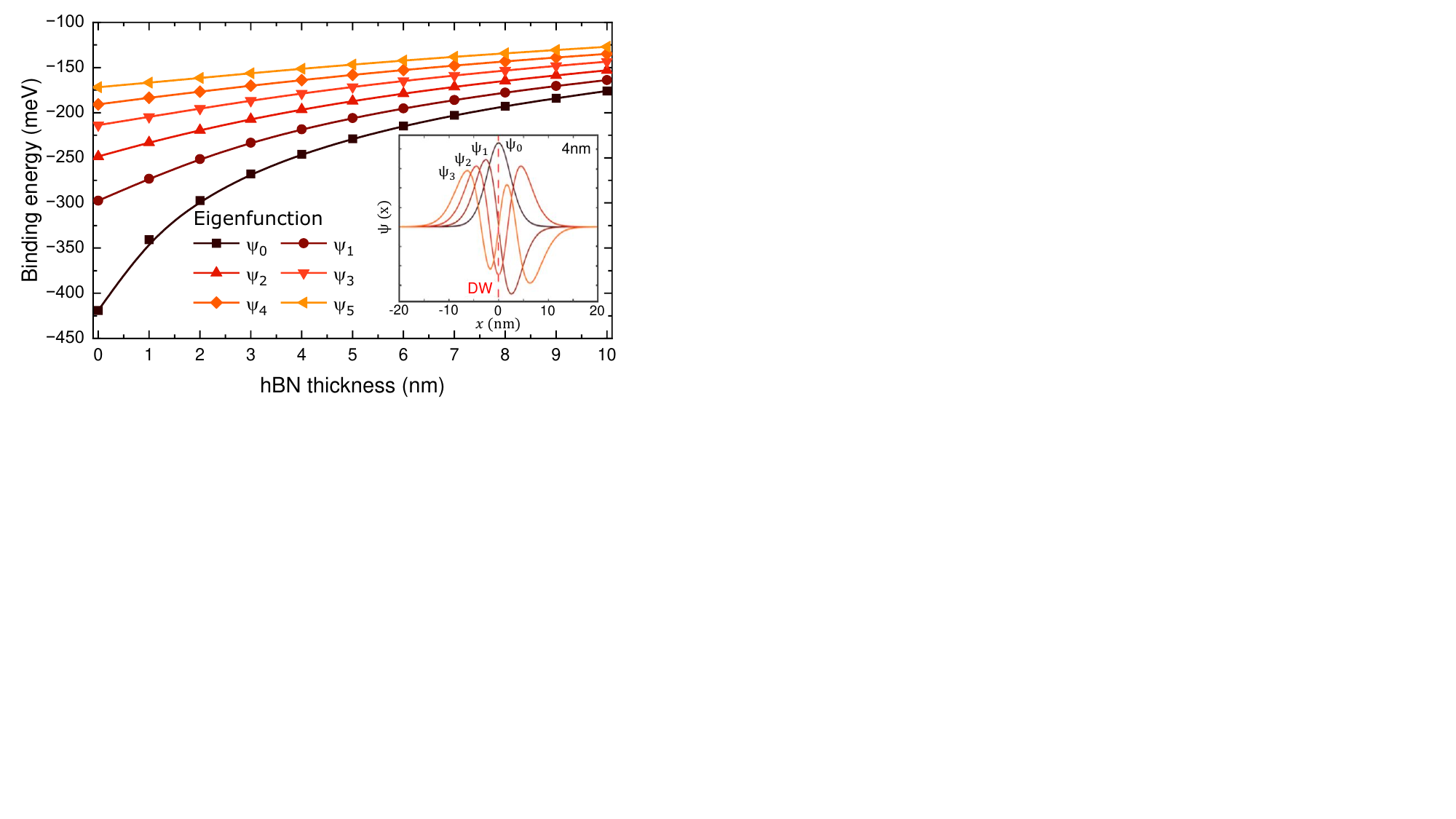}
\caption{\textbf{Potential trap and confined states.} Theoretical calculation of the eigenenergies of the confined states as a function of the bottom hBN thickness. Inset: First wavefunctions for a sample stacked on 4\,nm hBN.}
\label{figura:SM3}
\end{figure}

Using an in-plane exciton polarizability for 1L-MoSe$_2$ of $\alpha=6.5$nm$^2$V$^{-2}$~\cite{cavalcante2018stark} and the calculated $E_x(x)$ (see Fig.~\ref{figura:SM1}c), we determined the theoretical potential $V_{Stark}(x)$ presented in figure \ref{figura:SM3}a. By numerically solving the Schr\"odinger equation for this DC Stark potential trap, we obtained the wavefunctions and eigenenergies of the system, assuming an exciton mass of 1.2895$m_0$, where $m_0$ is the electron mass~\cite{LI20231312}. The eigeneneries for the first confined states ($\psi_0$ to $\psi_5$) are plotted in figure \ref{figura:SM3}b as a function of the bottom hBN thickness. The inset in Fig.\ref{figura:SM3}b illustrates the wavefunctions for the first confined states for a sample stacked on top of 4\,nm bottom hBN.

%==FIGURA======FIGURA======FIGURA======FIGURA======FIGURA======FIGURA=====
%==FIGURA======FIGURA======FIGURA======FIGURA======FIGURA======FIGURA=====

\subsection{Electronic landscape effects over $LX$s}\label{SN4}

%==FIGURA======FIGURA======FIGURA======FIGURA======FIGURA======FIGURA=====
\begin{figure}[t!!]
\includegraphics*[keepaspectratio=true, clip=true, angle=0, width=1\columnwidth, trim={0mm, 0mm, 0, 0mm}]{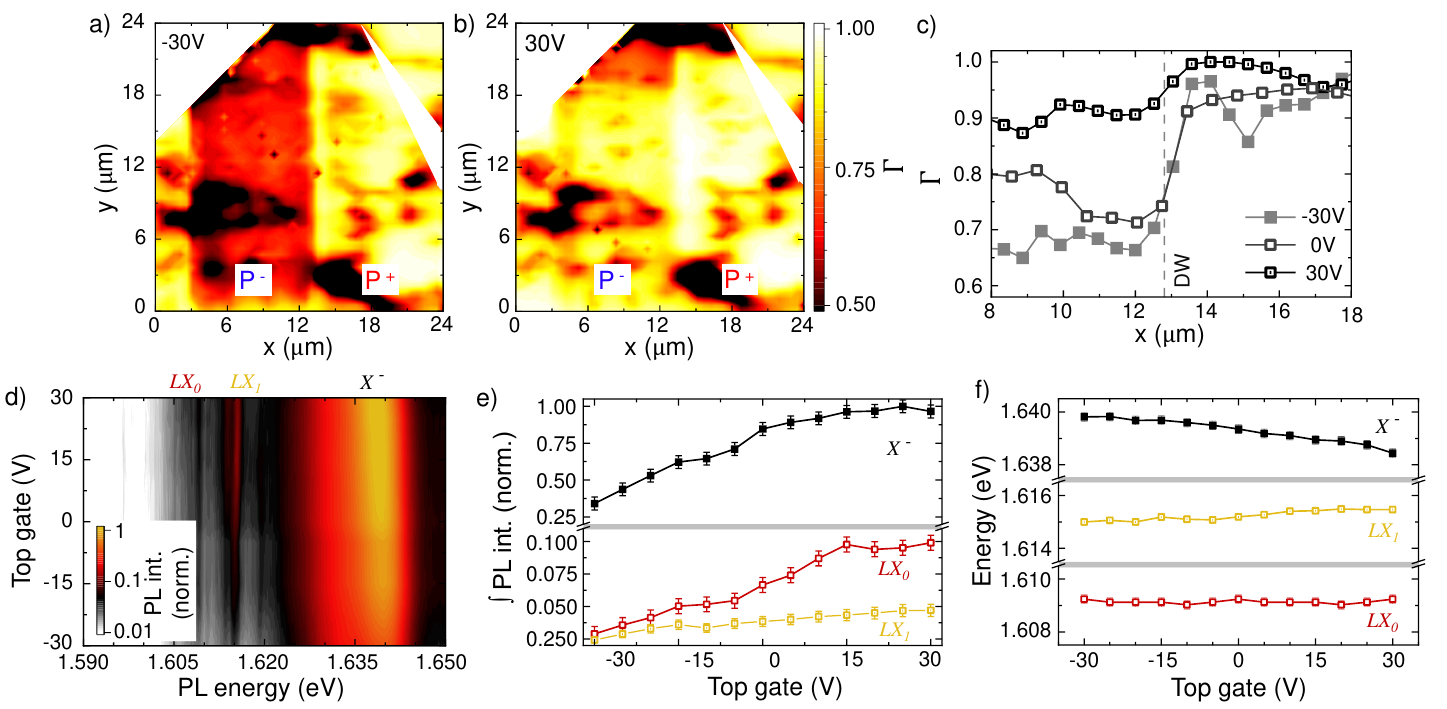}
\caption{\textbf{Electronic landscape effect over $LX$s. a} and \textbf{b)} False colour maps showing the top gate effect over the MoSe$_2$ photophysics through the parameter $\Gamma = I_{X^-}/(I_{X^-}+I_X)$. \textbf{c)} 
$\Gamma$ modulation across the DW for different top gate voltages. \textbf{d)} False colour plot of the $LX$ and trion PL spectra in resonance as a function of the top gate. \textbf{e)} $X^-$, $LX_0$ and $LX_1$ intensity as a function of the top gate voltage. \textbf{f)} $X^-$, $LX_0$ and $LX_1$ spectral position as a function of the top gate.}
\label{figura:SM4}
\end{figure}
%==FIGURA======FIGURA======FIGURA======FIGURA======FIGURA======FIGURA=====

while a comprehensive investigation of the material response lies beyond the scope of this work, we analyzed the influence of the free charge density on the $LX$ emission by tuning the sample top gate (sample 1). To characterize the effect of the top gate on the photophysics of the 2D counterpart, we introduce the parameter $\Gamma = I_{X^-}/(I_{X^-}+I_X)$. Figure \ref{figura:4}a and b present false-colour maps of $\Gamma$ across the sample under top gate voltage of -30\,V and 30\,V, respectively. At -30\,V, the $P^-$ and $P^+$ domains exhibit distinctly different charge density, reflected by $\Gamma\simeq0.65$  for the $P^-$ domain and $\Gamma\simeq0.95$ for the $P^+$ domain. In contrast, the map at 30\,V reveals a much more homogeneous $\Gamma$ across the sample. Note that as $\Gamma\rightarrow1$, the $X$ emission is suppressed. The effect of the top gate is summarized in Figure \ref{figura:4}c, which presents $\Gamma$ across the DW for three different voltages. Figure \ref{figura:4}d presents a false-colour plot of the in-resonance PL at the DW as function of the top gate voltage. Sweeping the top gate in the positive direction increases $X^-$ intensity and induces a redshift of 2\,meV, consistent with previous reports~\cite{ross2013electrical}. While the $X$ emission is suppressed by increasing the charge density at positive voltages, the $LX$s emission increases. Figure \ref{figura:4}e and f show the extracted intensity and spectral position for $LX_0$, $LX_1$ and $X^-$ as a function of the top gate voltage. While $LX$s and $X^-$ increase their intensity with similar proportion, the spectral position of $LX_0$ and $LX_1$ remains nearly constant, displaying a slight blueshift of $\sim$0.5\,meV.

%The lack of spectral shift in the $LX$ emission indicates that the depth and narrow profile of $V_{eff}$ are unaffected by the free charge density. This behaviour contrasts with similar studies in the literature~\cite{heithoff2024valley, thureja2022electrically}, where the effective potential, in addition to the Stark shift, is influenced by the interaction with repulsive polarons~\cite{efimkin2017many} forming an interaction-induced confinement. In our platform, the confinement potential is not a result of many-body interactions and is, therefore, insensitive to variations in local free charge density. Conversely, the increased $LX$ emission suggests changes in $LX$ dynamics. Given that $LX$ formation and radiative recombination are independent of free charge density, this may reflect a reduction in non-radiative channels, \hl{leading to an extended $LX$ lifetime. citations??}

\twocolumngrid

%
%###############################################################################
%								BIBLIOGRAPHY
%##############################################################################
%

\bibliographystyle{apsrev4-2}
\bibliography{bibliography}

\end{document}